\documentclass[reprint,amsmath,amssymb, aps]{revtex4-1}

\usepackage{graphicx}
\usepackage{dcolumn}
\usepackage{bm}

\begin{document}

\title{Magnetotransport properties of the new-type topological semimetal ZrTe}

\author{W. L. Zhu$^{1,3}$}
\thanks{These authors contributed equally to this work.}
\author{J. B. He$^{1,2}$}
\thanks{These authors contributed equally to this work.}
\author{S. Zhang$^{1}$}
\author{D. Chen$^{1}$}
\author{L. Shan$^{1,4}$}
\author{Z. A. Ren$^{1,4}$}
\author{G. F. Chen$^{1,3,4}$}
\email{gfchen@iphy.ac.cn}

\affiliation{
$^1$Institute of Physics and Beijing National Laboratory for Condensed Matter Physics, Chinese Academy of Sciences, Beijing 100190, China\\
$^2$College of Physics and Electronic Engineering, Nanyang Normal University, Nanyang 473061, China\\
$^3$School of Physical Sciences, University of Chinese Academy of Sciences, Beijing 100190, China\\
$^4$Collaborative Innovation Center of Quantum Matter, Beijing 100190, China\\
}
\date{\today}
\begin{abstract}

We report the first experimental results of the magnetoresistance, Hall effect, and quantum Shubnikov-de Haas oscillations on single crystals of ZrTe, which was recently predicted to be a new type of topological semimetal hosting both triply degenerate crossing points and Weyl fermion state. The analysis of Hall effect and quantum oscillations indicate that ZrTe is a multiband system with low carrier density, high carrier mobility, small cross-sectional area of Fermi surface, and light cyclotron effective mass, as observed in many topological semimetals. Meanwhile, the angular dependence of the magnetoresistance and the quantum-oscillation frequencies further suggest that ZrTe possesses a three-dimensional Fermi surface that is rather complex. Our results provide a new platform to realize exotic quantum phenomena related to the new three-component fermions distinct from Dirac and Weyl fermions.

\end{abstract}

\maketitle

\section*{INTRODUCTION}

The recent discovery of topological semimetals (TSMs) have sparked intense research interest in condensed matter physics and material science because their fascinating topological-protected band structures can lead to exotic physical phenomena and novel future applications \cite{TSM_W,TSM_RMP1,TSM_RMP2}. The TSMs are usually separated into several subfamilies according to the degeneracy, momentum space distribution, and the protecting mechanism of the band crossing points near Fermi level ($E_{F}$). For instance, in Dirac semimetals (DSMs) \cite{Na3Bi_C,Na3Bi_ARPES,Na3Bi_MR,Cd3As2_C,Cd3As2_ARPES,Cd3As2_MR} and Weyl semimetals (WSMs) \cite{TaAs_Science,TaAs_PRX_Weng,TaAs_PRX_Ding,TaAs_PRX_Chen}, four-fold degenerate Dirac point and two-fold degenerate Weyl point originate from the crossing of two double- and non-degeneracy bands, respectively. In the meanwhile, DSMs are protected by special crystalline symmetries on high symmetrical crystal momentum points or along high symmetry lines \cite{Na3Bi_C,Na3Bi_ARPES,Na3Bi_MR,Cd3As2_C,Cd3As2_ARPES,Cd3As2_MR}, and WSMs are stabilized without any additional crystalline symmetry besides the lattice translation \cite{TaAs_Science,TaAs_PRX_Weng,TaAs_PRX_Ding,TaAs_PRX_Chen}. Compared to DSMs and WSMs, in node-line semimetals (NLSMs), two bands cross in the form of a periodically continuous line or closed ring in the momentum space \cite{NLSM1,NLSM2}.

Very recently, some new types of TSMs identified by three-, six-, or eight-fold band crossings near the $E_{F}$ were proposed \cite{TPSM_SCience}. These so called ``New Fermions'' were predicated to possess novel physical properties which are far beyond the comprehension of DSMs and WSMs \cite{TPSM_SCience}. The three-fold degenerate crossing points (TPs) are formed by the crossing of a double-degeneracy band and a non-degeneracy band, and protected in a lattice either by rotation symmetries or nonsymmorphic symmetries \cite{TPSM_SCience}. The first-principle calculations indicated that the TPs might exist in the materials with WC-type structure, such as MoP, WC, TaN, and ZrTe \cite{TPSM_PRX,WC_Hasan,TaN_PRB,ZrTe_PRB}. Subsequently, the measurement of angle-resolved photoemission spectroscopy (ARPES) declared the presence of TPs in MoP \cite{MoP_ARPES}. However, MoP is thought to behave like a traditional metal with a little contribution from the three-component fermions to the transport properties due to that the TPs lie far below $E_{F}$ \cite{MoP_ARPES,MoP_MR}. While the recent ARPES experiments and transport measurements have shown that the TPs in WC lie much closer to $E_{F}$, which clearly results in unique transport behaviors, proving the nontrivial topological nature of the ``New Fermions" \cite{WC_ARPES,WC_MR}.

In this work, we have successfully grown the single crystals of WC-type ZrTe and performed the magnetotransport studies. Hall resistivity measurement indicates that ZrTe is a multiband system and the hole-type carriers with high mobility dominate the transport properties. The fast Fourier transform of the Shubnikov-de Haas (SdH) oscillations reveals four fundemental oscillation frequencies, arising from four small Fermi surface (FS) pockets with light cyclotron effective masses. Together with the angular dependence of the quantum-oscillation frequencies and the magnetoresistance [MR = $(\rho(B)-\rho(0))/\rho(0)$], our results indicate that ZrTe possesses a rather complex three-dimensional FS.

\section*{EXPERIMETNAL DETAILS}

The single crystals of WC-type ZrTe were grown from In-flux. Stoichiometric amounts of Zr and Te with moderate In were put into alumina crucible and sealed in an evacuated quartz tube. The quartz tube was heated to 1000$^\circ$C and then cooled slowly to 500$^\circ$C, where the In was decanted using a centrifuge. The obtained single crystals are hexagonal shape with sides of about 0.5 mm and thickness of 0.4 mm, as shown in the left inset of Fig. \ref{XRD}(b). The single crystals were characterized by x-ray diffraction (XRD) on a PANalytical diffractometer with Cu $K_{\alpha}$ radiation at room temperature. Magnetoresistance and Hall resistivity measurements were performed on a Quantum Design Physical Property Measurement System (PPMS). The average element compositions were determined by Oxford X-Max energy dispersive x-ray (EDX) spectroscopy analysis on a Hitachi S-4800 scanning electron microscope.

\section*{RESULTS AND DISCUSSION}

\begin{figure}
\includegraphics[width=8cm, height=7.5cm]{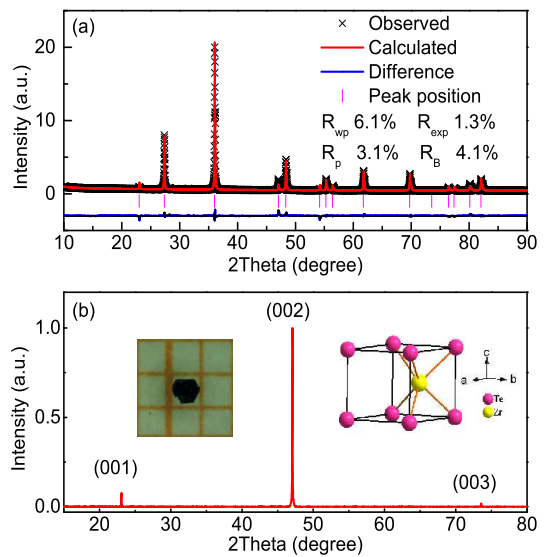}
\caption{\label{XRD}(Color online) (a) Powder XRD pattern of ZrTe obtained by grinding lots of single crystals at room temperature and the refinement results. (b) The single crystal XRD pattern of a WC-type ZrTe single crystal with (00\emph{l}) reflections. The left inset shows the optical image of a grown ZrTe single crystal sample. The right inset shows the crystal structure of WC-type ZrTe.
}
\end{figure}

Figure \ref{XRD}(a) shows the room-temperature powder XRD pattern of ZrTe obtained by grinding lots of single crystals. All the powder diffraction peaks can be indexed in a hexagonal WC-type crystal structure with space group \emph{P$\overline{6}m2$} (No.187) by the Rietveld refinement program, TOPAS-Academic \cite{XRD}. The refined lattice parameters $a = b = 3.764(1)$ {\AA} and $c = 3.860(1)$ {\AA} are in good agreement with the reported values \cite{ZrTe_XRD}. Figure \ref{XRD}(b) shows the reflection peaks of (00$l$) observed on a single crystal surface, suggesting that the crystal grows well along $ab$ plane as shown in the left inset. The average Zr : Te atomic ratio determined using the EDX is very close to 1 : 1 and no foreign element like In was detected within the limitation of instrument resolution.

 \begin{figure}
\includegraphics[width=9cm, height=8cm]{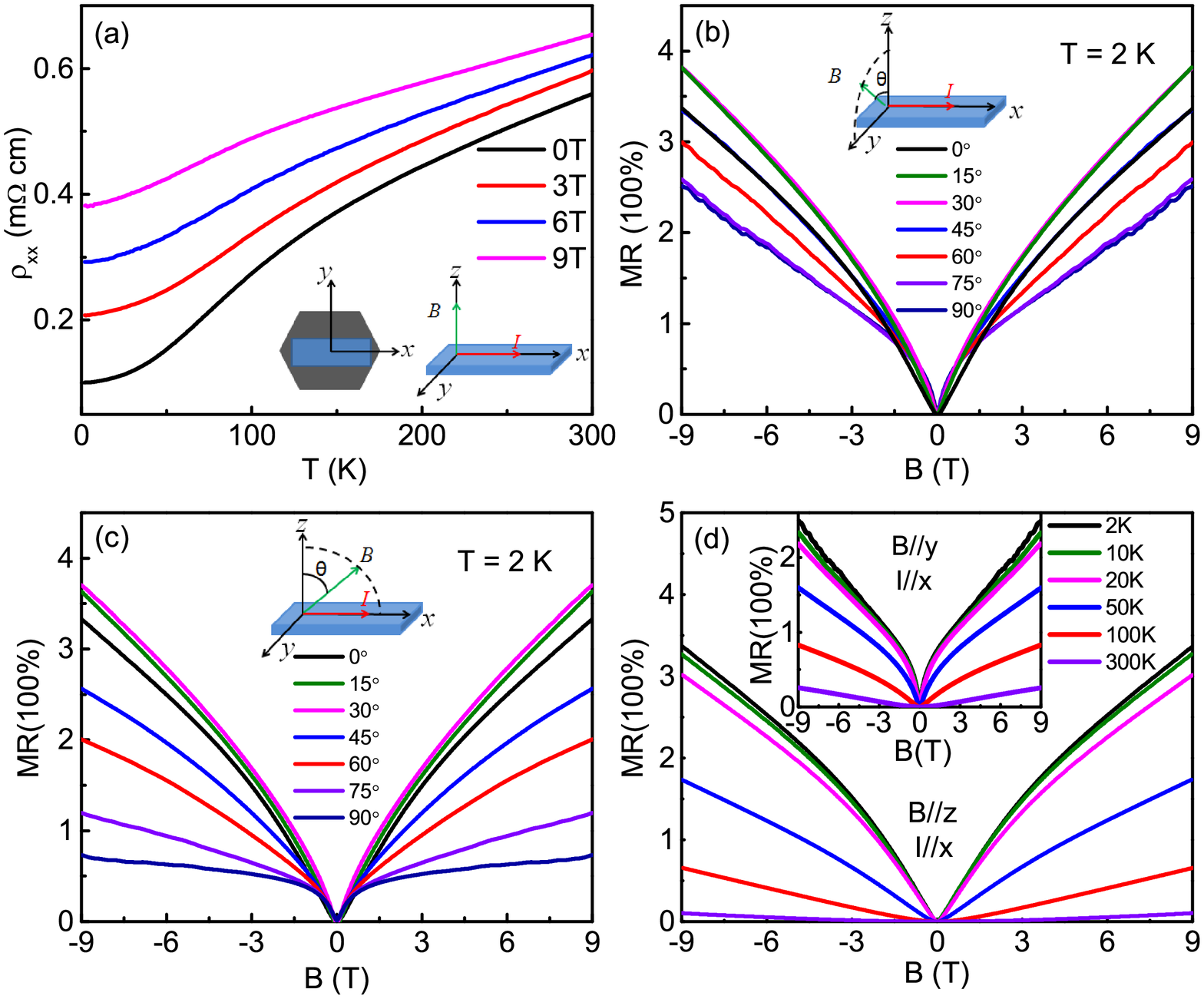}
\caption{\label{MR}(Color online) Transport properties of WC-type ZrTe. The direction of current parallel to the \emph{x}-axis (\emph{I} $\parallel$ \emph{x}). The insets of (a), (b), and (c) depict the measurement configurations. (a) Temperature dependence of the resistivity $\rho_{xx}$ under magnetic fields with \emph{B} $\parallel$ \emph{z}. (b) and (c) Magnetic field dependence of MR at 2 K with the magnetic field rotated in \emph{yz}- and \emph{xz}-plane, respectively. (d) Magnetic field dependence of MR at various temperatures for \emph{B} $\parallel$ \emph{z}. The inset shows the magnetic field dependence of MR at various temperatures for \emph{B} $\parallel$ \emph{y}.
}
\end{figure}

Figure \ref{MR}(a) shows the temperature dependence of the electrical resistivity $\rho_{xx}$(\emph{T}) measured in ZrTe under various magnetic fields with $B \parallel z$ and $I \parallel x$. In the case of zero field, $\rho_{xx}$(\emph{T}) shows a metallic behavior with a moderately large residual resistivity about 0.1 m$\Omega$ cm. With the application of magnetic fields, $\rho_{xx}$(\emph{T}) is enhanced. However, unlike the emergence of the field induced metal-to-insulator-like transition in other known DSMs and WSMs \cite{Na3Bi_MR,Cd3As2_MR,TaAs_PRX_Chen,WC_MR}, the resistivity of ZrTe lacks the low-temperature upturn even if the magnetic field up to 9 T. It will be very interesting to study the magnetotransport behaviors of ZrTe at much higher magnetic fields in the future. On the other hand, the lack of low-temperature upturn of $\rho_{xx}$(\emph{T}) under high magnetic fields lead directly to a small MR. In this case, the MR reaches only to 300 $\%$ for $B \parallel z$ and $I \parallel x$ at 2 K and 9 T.

We also noted that, at zero field, the residual resistivity ratio [RRR = $\rho_{xx}$(300 K)/$\rho_{xx}$(2 K)] is about $\sim$ 5, which is lower by one order of magnitude than that in the other TSMs \cite{Na3Bi_MR,Cd3As2_MR,TaAs_PRX_Chen,WC_MR}. It is found that the MR in TSMs scales well with the RRR due to different defect scatterings in the samples \cite{WTe2_RRR_MR}. Whereas in the case of ZrTe, the RRR appears to be largely independent of samples grown from different batches. The important point to note is that even for such low RRR of ZrTe, the quantum SdH oscillations can be clearly detected at a very low magnetic field (\emph{B} = 3 T). All of our data seem to suggest that low RRR and high residual resistivity are the intrinsic properties associated with band structure in ZrTe.

The isothermal magnetic field dependence MR at 2 K with rotating $B$ in the \emph{yz}-plane is shown in Fig. \ref{MR}(b). As scanning the $\theta$ from 0$^\circ$ to 90$^\circ$, the absolute value of MR increases firstly and then decreases, leading to a maximum near $\theta$ = 30$^\circ$ and a minimum at $\theta$ = 90$^\circ$, while the anisotropy of MR$_{\rm max}/$MR$_{\rm min}$ $\sim$ 1.5, is rather small. Here the non-monotonically angle-dependent and low anisotropic MR further suggests that ZrTe possesses a complex three-dimensional FS. In high magnetic fields, MR shows a linear-like field dependence, whereas MR decreases sharply with decreasing $B$ in low-field regime ($B<$ 1 T). This observation is reminiscent of the weak antilocalization (WAL) effect that has widely been reported in TSMs, such as TaAs and $\alpha$-As, in which the Dirac fermions dominate the transport process \cite{TaAs_PRX_Chen,As_PRB}. As shown in Fig. \ref{MR}(c), when we rotated the magnetic field in the \emph{xz}-plane, we also found that MR increases firstly at smaller titling angles and then decreases more quickly. However, no negative magnetoresistance was observed when the magnetic field is parallel to the current, which is consistent with the result in WC under the same measurement configuration \cite{WC_MR}. Figure \ref{MR}(d) shows the magnetic field dependence of MR at various temperatures for \emph{B} $\parallel$ \emph{z} and \emph{B} $\parallel$ \emph{y}. With increasing temperature, MR drops off obviously, and the WAL effect detected at low field gradually weakens for both directions.

\begin{figure}
\includegraphics[width=9cm, height=7cm]{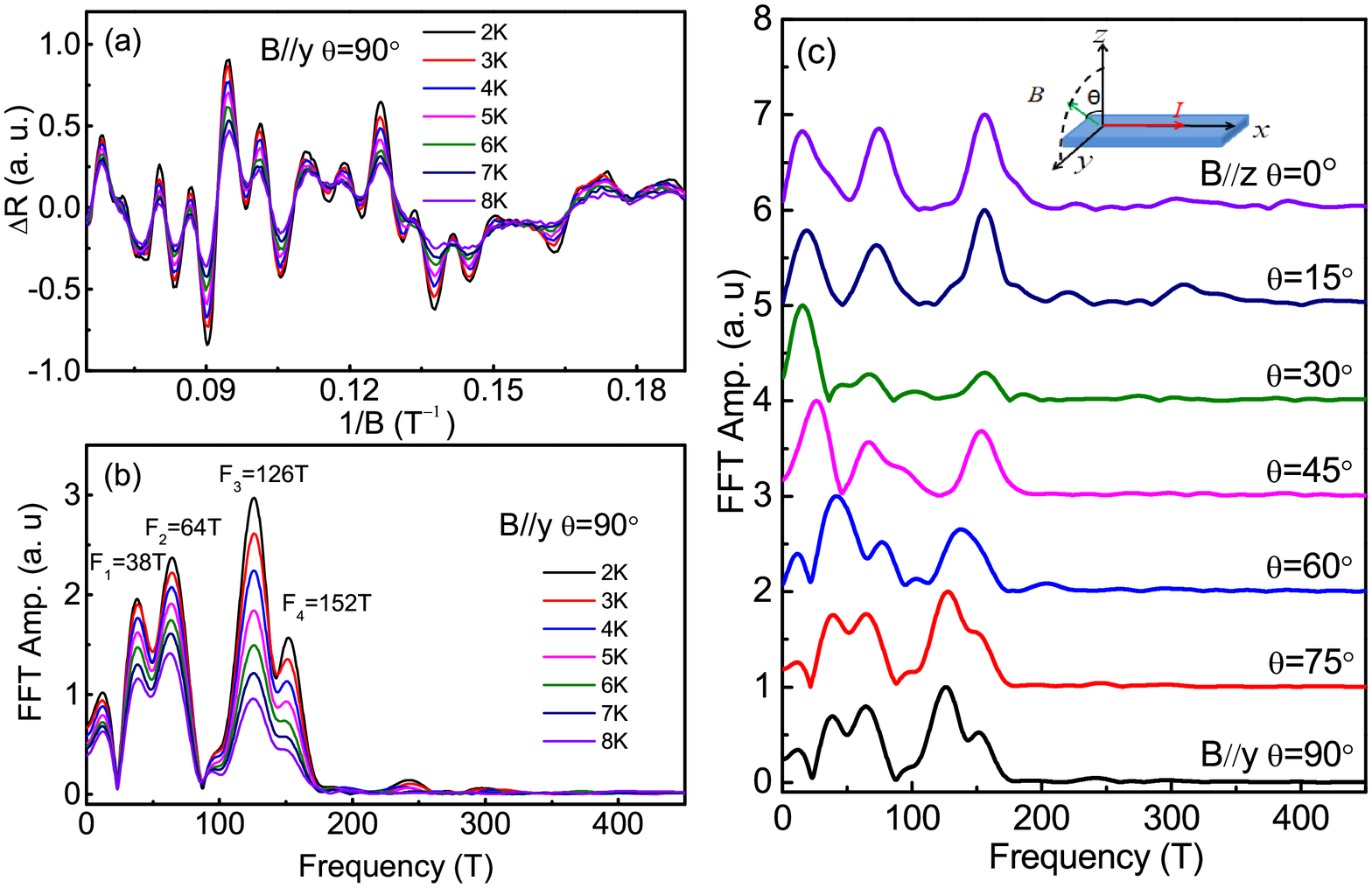}
\caption{\label{SDH}(Color online)(a) SdH oscillations after subtracting background at various temperatures for \emph{B} $\parallel$ \emph{y} ($\theta$ = 90$^\circ$). (b) The FFT spectra of the corresponding SdH oscillations with four fundamental frequencies at various temperatures for \emph{B} $\parallel$ \emph{y} ($\theta$ = 90$^\circ$). (c) Angular dependence of the FFT spectra of the corresponding SdH oscillations taken at 2 K by rotating the field from \emph{B} $\parallel$ \emph{z} ($\theta$ = 0$^\circ$) to \emph{B} $\parallel$ \emph{y} ($\theta$ = 90$^\circ$). The inset
depicts the measurement configuration.
}
\end{figure}

Quantum SdH oscillations of the resistance is one of the well established tools to probe the characteristics of FS. In order to obtain more reliable information from the SdH oscillations, we performed the MR measurements at the temperatures below 10 K in the magnetic field from 6 to 16 T. As shown in Fig. \ref{SDH}(a), we present the $1/B$ dependence of oscillatory components $\Delta R$ at several temperatures obtained by subtracting a background from the $R(B)$ isotherms. The oscillatory spectra are more complex, indicating the existence of multiple FS in this system. This can be seen more clearly by the fast Fourier transform (FFT) analysis as shown in Fig. \ref{SDH}(b). Four fundamental frequencies at about \emph{F}$_{1}$ = 38 T, \emph{F}$_{2}$ = 64 T, \emph{F}$_{3}$ = 126 T, and \emph{F}$_{4}$ = 152 T, and associated harmonics are observed. Although the frequency $F_3$ is almost the same as the 2nd harmonic of $F_2$, the fact that the FFT amplitude of $F_3$ is clearly larger than that of $F_2$ excludes this possibility. By using the Onsager relation $F = (\Phi_{0}/2\pi^{2})A_{F}$, where $A_{F}$ is the cross-sectional area of FS, \emph{F} is the FFT frequency, and $\Phi_{0}$ is the flux quantum, the cross-sectional area of the FS can be roughly estimated. The corresponding $A_{F}$ of the four frequencies are 0.36, 0.61, 1.23, and 1.54 nm$^{-2}$, respectively, which only occupy $0.11\%$, $0.19\%$, $0.38\%$, and $0.48\%$ of the first Brillouin zone. The cyclotron effective masses $m^{\ast}$ = 0.16, 0.15, 0.24, and 0.26 $m_{e}$ ($m_{e}$ is the bare electronic mass) for the four frequencies are obtained by fitting the temperature dependence of FFT amplitude ($Amp.$) with the following Lifshitz-Kosevitch formula \cite{SdH}:

\begin{equation}\label{1}
\emph{Amp.} \propto \frac{2\pi^{2}k_{B}Tm^{\ast}/{\hbar}eB}{sinh(2\pi^{2}k_{B}Tm^{\ast}/{\hbar}eB)},
\end{equation}
where $\hbar$ is Plancks constant divided by 2$\pi$, $k_{B}$ is the Boltzmann constant, \emph{e} is the bare electron charge, and $m^{\ast}$ is cyclotron effective mass. The light cyclotron effective mass and the small cross-sectional area of FS are comparable to those observed in some other TSMs such as TaAs and WC \cite{TaAs_PRX_Chen,WC_MR}. Figure \ref{SDH}(c) shows the FFT spectra of the SdH oscillations at 2 K obtained upon rotating the field from \emph{B} $\parallel$ \emph{z} ($\theta$ = 0$^\circ$) to \emph{B} $\parallel$ \emph{y} ($\theta$ = 90$^\circ$) in the \emph{yz-}plane. The four frequencies show rather weak change with rotating $\theta$ from 90$^\circ$ to $60^\circ$ and from 30$^\circ$ to 0$^\circ$, whereas a clear shift of the frequency occurs between 60$^\circ$ and 30$^\circ$, implying again the rather complicated FS character in the WC-type ZrTe.

 \begin{figure}
\includegraphics[width=9cm, height=7cm]{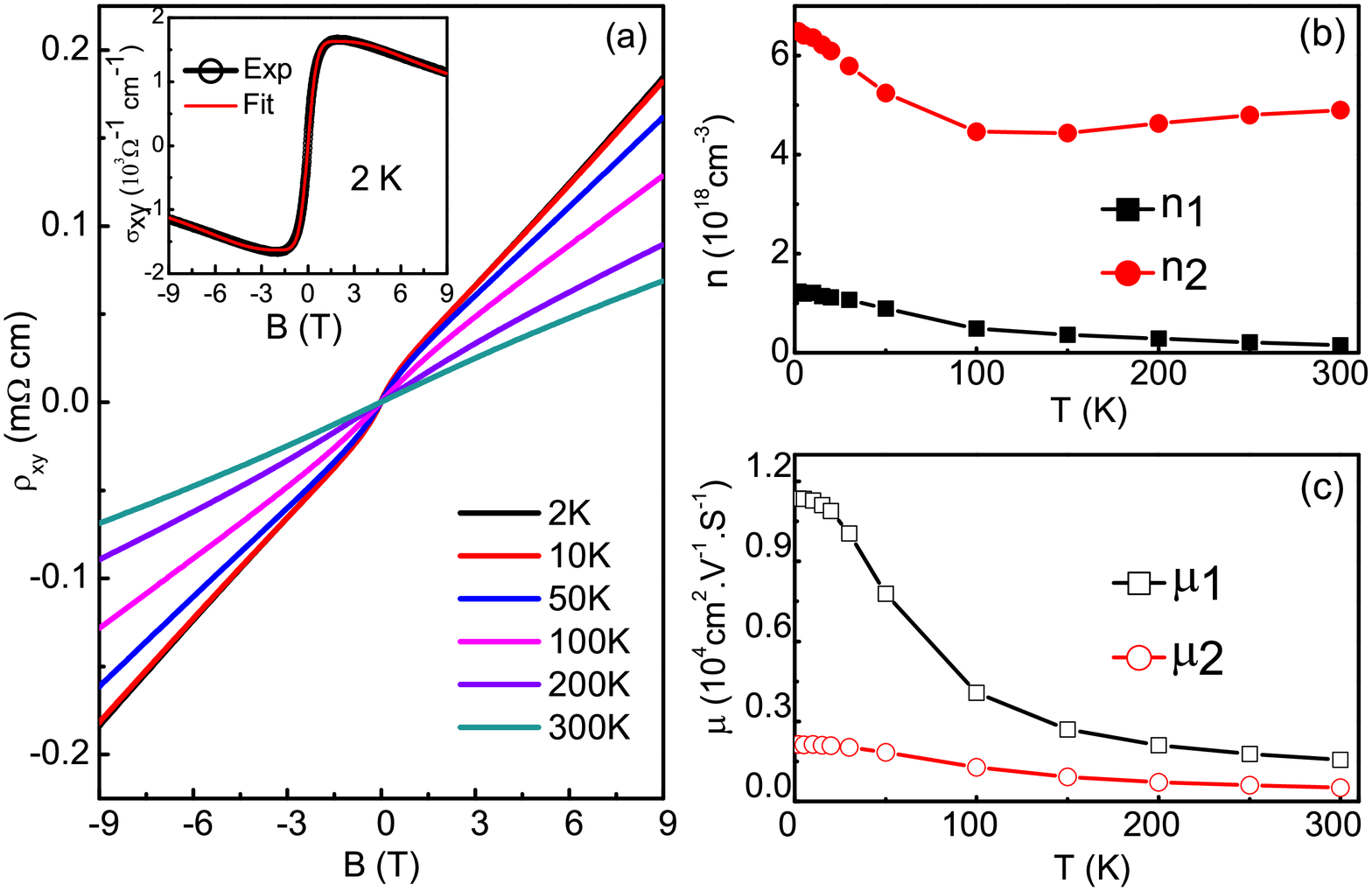}
\caption{\label{Hall}(Color online) (a) The magnetic field dependence of Hall resistivity $\rho_{xy}$ at selected temperatures. The inset shows the experimental data of Hall conductivity at 2 K and the fitted curve. (b) and (c) The temperature dependence of carrier densities and carrier mobilities obtained by fitting Hall conductivity $\sigma_{xy}$ using the two band model.
}
\end{figure}

The multiband character of ZrTe is further confirmed by carrying out Hall resistivity $\rho_{xy}$ measurements. As shown in Fig. \ref{Hall}(a), $\rho_{xy}$ shows a linear field dependence at the fields higher than 2 T, whereas a nonlinear field dependent $\rho_{xy}$ presents at $B<1$ T. The deviation from the linear dependence of $\rho_{xy}$ indicates clearly that ZrTe is a multiband system. Here, we fitted the Hall conductivity $\sigma_{xy}$ using a semiclassical two-band model \cite{Hall_1st,Hall_2nd}:
\begin{equation}\label{2}
\sigma_{xy} = [\frac{n_{1}\mu_{1}^{2}}{1+(\mu_{1}B)^2}+\frac{n_{2}\mu_{2}^{2}}{1+(\mu_{2}B)^2}]eB,
\end{equation}
where $n_{1}$ (or $n_{2}$) and $\mu_{1}$ (or $\mu_{2}$) are the carrier density and mobility for two kinds of carriers, respectively. We found that the $\sigma_{xy}$ between 2 and 300 K can be well fitted using the two-band model, and the inset of Fig. \ref{Hall}(a) shows a typical fitting curve for $T$ = 2 K. Our results strongly suggest that at least two channels of hole-like carriers are responsible for the observed transport behavior in ZrTe, and the calculated carrier density is in the order of $10^{-18}$ cm$^{-3}$, as shown in Fig. \ref{Hall}(b). One can clearly see that the two kinds of Hall mobility exhibit distinct behaviors with respect to temperatures, as depicted in Fig. \ref{Hall}(c). $\mu_{2}$ has a weak dependence on temperature. However, with decreasing temperature especially below 100 K, $\mu_{1}$ increases dramatically and reaches up to 1.14 $\times$ 10$^{4}$ cm$^{2}$ V$^{-1}$ s$^{-1}$ at 2 K, which is larger than $\mu_{2}$ by one order of magnitude. The low carrier density and high mobility, both of which are the key features of Dirac and Weyl fermions \cite{Cd3As2_MR,TaAs_PRX_Chen,WC_MR}. But there is much difference from most other TSMs \cite{WC_MR,As_PRB}, in which the presence of both electron- and hole-pockets with similar volumes provides partial charge compensation that is responsible for their giant magnetoresistance.

\section*{CONCLUSION}

In conclusion, we have performed the magnetotransport measurements on the single crystal of WC-type ZrTe, a proposed new-type of TSMs with triply degenerate points. Unlike in the well-studied DSMs and WSMs, the low-temperature resistivity does not show a field induced metal-to-insulator-like transition up to 9 T. Quantum SdH oscillations are visible in MR under fields above 3 T and are composed of a mixture of multiple frequencies, arising from the small Fermi surface pockets with light cyclotron effective masses. The angular dependence of the oscillation frequencies and the MR indicate that ZrTe consists of a rather complex three-dimensional Fermi surface. MR at low temperatures shows a weak WAL effect that depends strongly on the field direction. The Hall effect illustrates that ZrTe is a multiband system with low carrier density and high carrier mobility. Remarkably, unlike in most other TSMs, the same type of charge carriers but with different values of mobility are responsible for the observed transport behavior in ZrTe.

\section*{ACKNOWLEDGEMENTS}

This work was supported by the National Basic Research Program of China 973 Program (Grant No. 2015CB921303), the National Key Research Program of China (Grant No. 2016YFA0300604), the Strategic Priority Research Program (B) of Chinese Academy of Sciences (Grant No. XDB07020100), and the Natural Science Foundation of China (Grant No. 11404175).

\bibliography{References}

\end{document}